\documentclass[useAMS,usenatbib]{mn2e}
\usepackage{amssymb}
\usepackage{amsmath}
\usepackage{txfonts}
\usepackage{graphicx}
\usepackage{natbib}
\usepackage{rotating}
\usepackage{url}
\usepackage{epsfig}
\usepackage{color}
\usepackage{supertabular}
\usepackage{array}

\newcommand{\swift}{{\it Swift}}
\newcommand{\gr}{$\gamma$-ray}

\renewcommand{\theequation}{\thesection.\arabic{equation}}

\begin{document}

\date{\empty}

\title{Where are {\it Swift} $\gamma$-ray bursts beyond the "synchrotron deathline"?}
   \author[Savchenko and Neronov]{V.Savchenko$^{1,2}$,
          A.Neronov$^{1,2}$\\
             $^1$ISDC Data Centre for Astrophysics, Ch. d'Ecogia 16, 1290, Versoix, Switzerland\\
  $^2$Geneva Observatory, Ch. des Maillettes 51, 1290, Sauverny, Switzerland       }

\date{Received $<$date$>$  ; in original form  $<$date$>$ }

\pagerange{\pageref{firstpage}--\pageref{lastpage}} \pubyear{2007}

\maketitle

\label{firstpage}

\begin{abstract}
We study time-resolved spectra of the prompt emission of \swift\ $\gamma$-ray bursts (GRB). Our goal is to see if previous BATSE claims of the existence of a large amount of spectra with the low energy photon indices harder than $2/3$ are consistent with \swift\ data. We perform a systematic search of the episodes of the spectral hardening down to the photon indices $\le 2/3$ in the prompt emission spectra of \swift\ GRBs.  We show that the data of the BAT instrument on board of \swift\ are consistent with BATSE data, if one takes into account differences between the two instruments. Much lower statistics of the very hard spectra in \swift\  GRBs is explained by the smaller field of view and narrower energy band of the BAT telescope. 
\end{abstract}

\section{ Introduction }\label{intro}

In spite of the fact that the phenomenon of the $\gamma$-ray bursts (GRB) was discovered more than 40 years ago, its origin and basic physical mechanisms of the observed  0.1-100 second long pulses of hard X-ray /soft $\gamma$-ray emission remain obscure.  A significant progress in understanding of the GRB phenomenon was achieved over the last decade with the discovery of X-ray, optical and radio afterglows of long (duration $>2$~s) \citep{vanPar97,Frail97} and short (duration $\le 2$~s) GRBs \citep{Gehrels05,Barthelmy05,Fox05,Hjorth05}. Localization of long GRBs in star formation regions, where more than 90\% of the supernovae occur, indicates that long GRBs may be produced by the death of massive stars in supernova explosions. This is confirmed by the direct observation of appearance of supernovae  at the GRB positions \citep{Galama98,Stanek03}.

The mechanism of production of GRBs is constrained not only by the identification of their multi-wavelength counterparts, but also by the intrinsic properties of the \gr\ emission. Observations of fast, millisecond-scale, variability during the prompt emission phase \citep{bhat92,walker00} point to the  presence of a compact "central engine" powering the GRB, which is naturally associated with a stellar mass black hole or a neutron star formed in result of the gravitational collapse of a massive star at the on-set of a supernova explosion. \gr s can escape from a relatively compact emission region only if the emitting medium moves with a large bulk Lorentz factor $\gamma_b > 100$ \citep{Fenimore93,WL95}. 

Although it is clear that the observed emission is produced by relativistically moving plasma ejected by a newly formed black hole or neutron star, the physical mechanism of the \gr\ emission is not well constrained by the available data.  One possibility is that this emission is synchrotron emission from electrons accelerated on "internal" shocks formed in collisions of plasma blobs moving with different velocities \citep{ZM04,Piran05}.  Alternatively, prompt \gr\ emission can be produced via inverse Compton scattering of lower energy synchrotron photons originating from the relativistic plasma itself (the so-called synchrotron-self-compton [SSC] model) \citep{panaitescu00,kumar,piran08}  or of the photons from external ambient radiation fields (external Compton [EC] model)  \citep{shemi94,shaviv95,lazatti00,DR04,piran08}. 
 
 Recent detections of strong prompt optical emission from several  GRBs seem to indicate the presence of a separate lower energy prompt emission component of the GRB emission \citep{akerlof99,verstrand05,verstrand06,racusin08}. A natural interpretation of this observation would be that optical and soft \gr\ emission are produced, respectively, via synchrotron and inverse Compton  mechanisms by one and the same population of relativistic electrons, thus favoring the inverse Compton (SSC or EC) scenario for the 0.01-10~MeV band emission. It is, however, possible that optical and \gr\ components of prompt emission are produced by two separate electron populations and/or in different emission regions \citep{piran08b}. A strong test of the inverse Compton model of the prompt \gr\ emission can be given in the nearest future via (non)detection of the predicted second-order inverse Compton scattering component in the 1-100~GeV energy band by {\it Fermi/GLAST} and/or ground-based \gr\ telescopes~\citep{racusin08,savchenko08}.  
 
The synchrotron and inverse Compton models of the prompt \gr\ emission could be distinguished  via a study of the spectral characteristics of the prompt GRB emission. If the prompt emission is optically thin synchrotron emission from the shock-accelerated electrons in a relativistic "fireball", the spectrum of \gr\ emission could not have photon index harder than the one of the synchrotron emission from a monoenergetic electron distribution, $\Gamma_{\rm synch,lim}=2/3$ \citep{Katz94,Tavani94}. At the same time, the spectrum of inverse Compton emission can be as hard as $\Gamma_{\rm IC,lim}=0$ \citep{BB04}. Observation of episodes of hardening of GRB spectra beyond $\Gamma_{\rm synch,lim}$ would provide a strong argument against the synchrotron model of the prompt emission (see, however, \citep{epstein73,lloyd00,lloyd02,medvedev00,BB04} for modifications of the synchrotron model which are able to accommodate photon indices harder than $2/3$).

The time-resolved spectral characteristics of the prompt GRB emission were studied in details by \citet{preece98a,preece98b,preece00} and \citet{kaneko08} using a set of bright GRBs detected by the BATSE instrument on board of {\it CGRO}.  The study of BATSE GRBs shows that approximately $30
\%$ of all the spectra violate the "synchrotron deathline" $\Gamma=2/3$  \citep{preece98b}. This apparently  rules out the synchrotron model as a viable model of the prompt \gr\ emission. 

In the view of  significance of this result, an independent test of the result with a different instrument  is important. However, no such independent test was available so far. In fact, the validity of BATSE result was questioned by the non-observation of excessively hard GRB spectra by {\it HETE} \citep{heteII}. The HETE-BATSE inconsistency could be resolved by observations with higher sensitivity instruments, such as \swift. 

In what follows we perform a systematic analysis of the time resolved spectra of \swift\ GRBs. We show that, contrary to the initial expectations, \swift\ provides only a very limited possibility for testing the BATSE claim of existence of a significant number of GRBs beyond the "synchrotron deathline". The main problem is that a proper reconstruction of the photon index in the GRB prompt emission spectrum requires a measurement of the break, or cut-off energy, which is not possible with \swift, because of the limited energy range of its Burst Alert Telescope (BAT). Another difficulty lies in the smaller field of view of \swift/BAT, compared to BATSE, which leads to a lower rate of detection of sufficiently bright GRBs, for which the quality of reconstruction of spectral parameters in the time-resolved spectra is comparable to the one of the sample of BATSE GRB spectra. Taking into account these difficulties, we show that \swift\ results on detections of very hard GRBs with photon indices beyond the "synchrotron deathline" are consistent with the expectations based on the extrapolation of BATSE results to the \swift\ energy band and sensitivity.  
 
The paper is organized as follows. In section \ref{simulations} we explore the potential of \swift\  to test the BATSE result on very hard GRB prompt emission spectra. To do this we simulate the appearance of BATSE GRBs in BAT telescope on board of \swift. We find that only partial tests of BATSE result are possible with \swift. In Section \ref{simul1} we precise this statement and formulate a quantitative prediction on the number of GRB spectra beyond the "synchrotron deathline" which is expected in the \swift\ data. Next, in sections \ref{sec:selection}, \ref{method} and \ref{res} we perform a systematic search of the hard prompt emission spectra in \swift\ GRBs. Finally, in Section \ref{discuss}  we discuss the implications of the observation of a limited amount of GRB spectra beyond the "synchrotron deathline" in \swift\ GRB sample.

\section{Spectra of prompt $\gamma$-ray emission of \swift\ GRBs}

The spectra of prompt emission from GRBs are conventionally modeled
with the "Band function" \citep{band93}
\begin{equation}
\label{band}
\frac{dN_\gamma}{dE}=A\left\{
\begin{array}{ll}
\left[\frac{\displaystyle E}{\displaystyle 50\mbox{
keV}}\right]^\alpha \exp\left(-\frac{\displaystyle
(\alpha-\beta)E}{\displaystyle E_{\rm break}}\right);& E<E_{\rm
break}
\nonumber\\
\left[\frac{\displaystyle E_{\rm break}}{\displaystyle 50\mbox{
keV}}\right]^{\alpha-\beta}
 \exp\left(\beta-\alpha\right)
 \left[\frac{\displaystyle E}{\displaystyle 50\mbox{ keV}}\right]^{\beta}; &E\ge E_{\rm
break}
\end{array}
\right.
\end{equation}
where $E_{\rm break}$ is the energy of the break in the spectrum, $\alpha$ and $\beta$ are, respectively, low and high energy photon (spectral) indices and $A$ is the normalization constant. The break energy is related  in a simple way to the peak energy of the spectral energy distribution (SED) of the GRB,  $E_{\rm peak}=(2+\alpha)E_{\rm break}/(\alpha-\beta)$. 

The BAT telescope on board of \swift\ is not optimized for the measurement of all the parameters of the Band model ($E_{\rm break}, \alpha, \beta$). It is sensitive in the energy range $15-150$~keV, while for most of the GRBs $E_{peak}$ is above $150$~keV. Instead, the BAT instrument seems to be well suited to the study of the low-energy part of the GRB spectrum, in particular, to the measurement of the low-energy photon index $\alpha$, which is the primary subject of our investigation. 

The low-energy part of the GRB spectrum is well described by a cut-off power-law function
\begin{equation}
\label{cutoffpl}
\frac{dN_\gamma}{dE}=A\left[\frac{E}{50\mbox{ keV}}\right]^{-\Gamma}\exp\left(-\frac{E}{E_{\rm cut}}\right)
\end{equation}
The photon index $\Gamma$ and the cut-off energy $E_{\rm cut}$ of the cut-off powerlaw model are simply related to the parameters of the Band model, 
\begin{equation}
\label{alpha_Ecut}
\Gamma=-\alpha;\ \ \ E_{\rm cut}=\frac{E_{\rm peak}}{(2+\alpha)}.
\end{equation}
Taking into account this one-to-one relation between the Band and cut-off powerlaw models in the \swift\ energy band, we use the cutoff powerlaw, rather than the Band, model for fitting the real \swift\ and simulated BATSE prompt GRB spectra in the following sections. 

In the cases when the cut-off energy $E_{\rm cut}$ is not constrained by the data (this can be either when $E_{\rm cut}>150$~keV or when $E_{\rm cut}\le 150$~keV, but the signal statistics at the highest energies  is not sufficient for the measurement of $E_{\rm cut}$) we further simplify the model spectrum and use a simple powerlaw model 
\begin{equation}
\frac{dN_\gamma}{dE}=A\left[\frac{E}{50\mbox{ keV}}\right]^{-\Gamma}
\end{equation}
for fitting of the real and simulated prompt emission spectra.

\section{How would BATSE GRBs look like in \swift/BAT?}\label{simulations}

Since the BAT instrument is well suited to the study of the low-energy part of the GRB spectrum, the  BATSE observation that some 30\% of the time-resolved GRB spectra have low-energy spectral indices harder than $\Gamma=2/3$ \citep{preece98a} should be readily testable with  \swift.  In order to see if this expectation is true, we first try to find out how the time-resolved spectra of BATSE GRBs, from the database collected by \citet{kaneko08}  would look like if they would be observed by \swift.

\begin{figure*}
\includegraphics[width=1.0\columnwidth,angle=0]{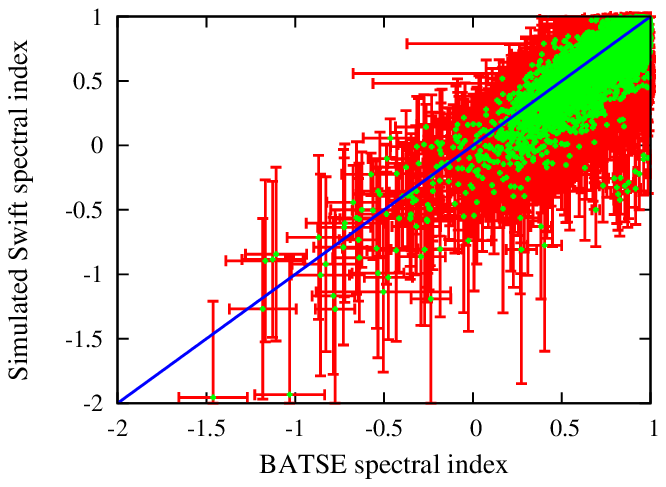}
\includegraphics[width=1.0\columnwidth,angle=0]{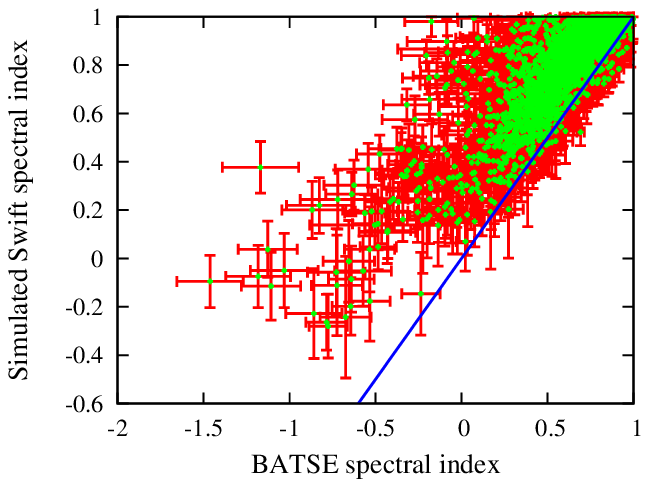}
\caption{Photon indices, found by fitting simulated BATSE spectra with the {\tt cutep50} (left panel) and with the {\tt power 50} (right panel) models of XSPEC vs the model photon indices. The model parameters are taken from the time-resolved BATSE spectra fitted on a cutoff powerlaw model. Solid line corresponds to the case when the reconstructed photon index is equal to the input model photon index.
}
\label{simul}
\end{figure*}

The BATSE Gamma Ray Burst catalog \citep{kaneko08} includes information on the spectral parameters of the time-resolved GRB spectra as well as information about the time interval durations\footnote{http://www.batse.msfc.nasa.gov/~kaneko/}. The entire database contains $\sim 8000$ time-resolved spectra from $\sim 300$ GRBs detected during $\sim 10$~years of operation of {\it CGRO} mission. To simulate the appearance of BATSE GRBs in \swift/BAT, we first extrapolate the reported GRB spectra to the BAT energy band (15-150~keV) and then simulate the BAT spectrum, using the XSPEC command {\tt fakeit}. We use the  {\tt batphasimerr} tool to calculate errors of the simulated spectra, as it is as described in the BAT analysis manual\footnote{http://heasarc.nasa.gov/docs/swift/analysis/threads/batsimspectrumthread.html}. In our simulations, we have found that the simulation procedure, described in the BAT analysis manual, does not take into account systematic errors, which are applied in normal data reduction procedure separately, using the  {\tt batphasyerr} tool. Taking this into account, we separately add a systematic error  to the simulated spectra, using {\tt batphasyerr} command. In principle, systematic errors, added by the {\tt batphasyerr} command, appear to depend on the observation date. We set all simulated spectra at a fake date,  the date of GRB060614.

The BAT telescope on board of \swift\ is a coded mask instrument. The sensitivity of a coded mask instrument is flat throughout the so-called "fully-coded" field of view (part of the field of view in which a source illuminates the entire detector), while it degrades in the partially coded part of the field of view (a part in which a source illuminates only a part of the detector). In our simulations we assume that the GRB always appears in the fully coded field of view. In principle, in a realistic situation it is not  true, especially at the initial stage of the GRB, when the BAT first triggers on a burst in a partially coded field of view and then slews toward the direction of the GRB. We discuss the implications of the neglect of the effects of the partially coded field of view in more details in the following sections. 

The results of simulations of BATSE GRBs are presented in Fig. \ref{simul}. From this figure one can derive several important implications for the search of the hard GRB spectra with \swift/BAT telescope. 
\begin{enumerate}
\item The errorbars of the spectral indices of the simulated BAT spectra are larger than the ones of the initial BATSE spectra, if both BAT and BATSE spectra are fit on one and the same model, the cut-off powerlaw model  (see left panel of Fig. \ref{simul}). This is explained by the fact that, in spite of the comparable signal statistics in the original and simulated spectra, the derivation of the errors of the measured spectral indices are affected by the uncertainty of the measurement of the cut-off energy $E_{\rm cut}$  in the BAT spectrum.  This is explained by the limited energy range of BAT detector.  
\item Taking into account the uncertainty of $E_{\rm cut}$ we have fitted the simulated spectra on the simple powerlaw model. This, obviously, reduces the error of the measurement of the photon index, making it comparable to the one of the initial BATSE spectra (right panel of Fig. \ref{simul}). However, ignoring the presence of the cut-off in the spectra results in a systematically softer reconstructed photon index in the simulated spectra. This significantly reduces the number of the very hard spectra, as compared to the original BATSE sample.
\end{enumerate}

\begin{table*}
\begin{tabular}{lccclcc}
\hline Data sample & Spectral model & Total number of indices & Indices smaller than $2/3$ & significance& Indices harder than $2/3$ at $\ge 3\sigma$ level\\
\hline
BATSE & {\tt cutep50} & 8072 & $2080 \pm 611 $ & 3.4 & 809 in 138 GRBs \\
BATSE & {\tt power50} & 8129 & $607 \pm 101 $ & 6.0 & 342 in 66 GRBs\\
\hline
BATSE, scaled (1) & {\tt cutep50} & 278 & $71  \pm 113 $ & & 28 in 5 GRBs\\
BATSE, scaled (1) & {\tt power50} & 300 & $22 \pm 19 $ &  & 13 in 2.4 GRBs\\
\hline
BATSE, scaled(2) & {\tt cutep50} & 278 & $81 \pm 190$ &  & 2 in 1 \\
BATSE, scaled(2) & {\tt power50} & 300 & $24 \pm 40$ & & 3.4 in 1 \\
 \hline
BAT & {\tt cutep50} & 343 & $19  \pm 8 $ & 2.4 & 4 in 2 GRBs\\
BAT & {\tt power50} & 367 & $7  \pm 3 $ & 2.3 & 0 in 0 GRBs\\
\hline
\end{tabular}
\caption{Statistics of GRB spectral parameters in different samples. Two top rows show the data for the simulations of BATSE GRBs from the sample of \citet{kaneko08} assuming BAT response. The sample marked "BATSE, scaled (1)" is the re-scaling of the full sample of simulated BATSE spectra, taking into account smaller exposure of \swift. The sample marked "BATSE, scaled (2)" is a re-scaling of the full simulated BATSE sample for the smaller number of \swift\ GRBs, with additional assumption that the error of the reconstructed photon index is two times larger than the one implied via the simulation procedure discussed in the text. Last two rows show the statistics of the spectral parameters for the sample of real \swift\ GRBs selected following the criteria given in Section \ref{sec:selection}.} 
\label{tnumbbeyond}
\end{table*}

Simulations of the time-resolved GRB spectra enable to estimate the expected amount of hard spectra in the \swift\ GRB sample. 

\begin{figure}
\includegraphics[width=1.05\columnwidth,angle=0]{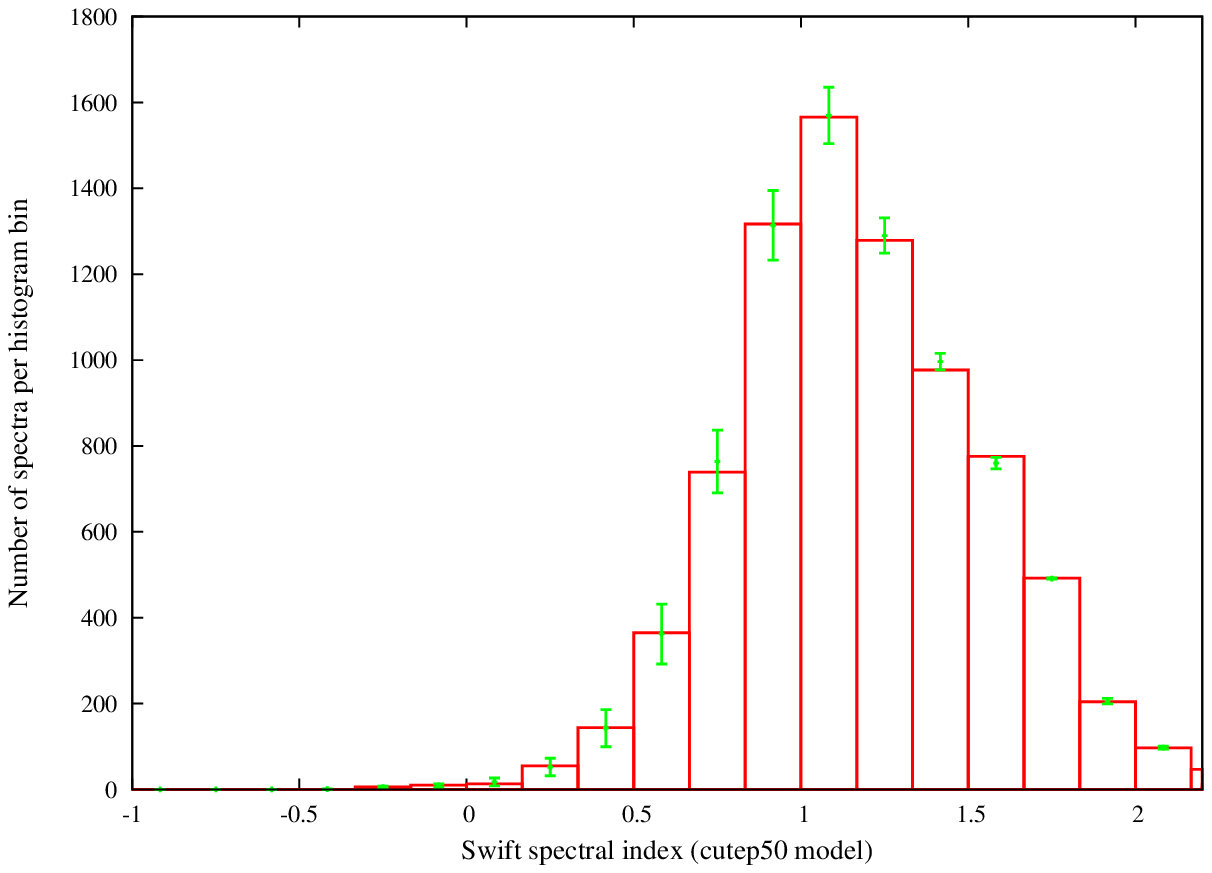}
\includegraphics[width=1.05\columnwidth,angle=0]{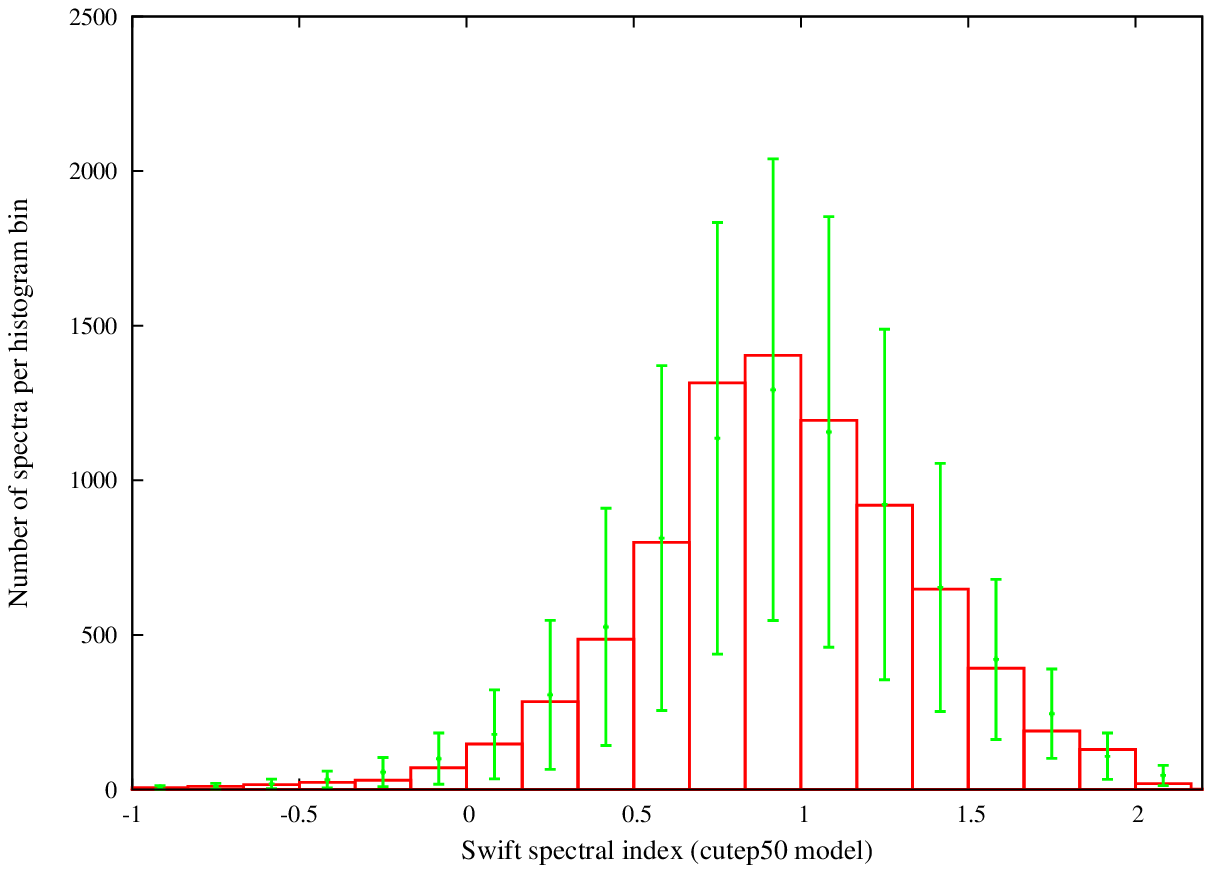}
\caption{Upper panel: histogram of the reconstructed low energy powerlaw indices (fit on the  {\tt power50} model) of simulated BATSE GRB spectra. Lower panel: the same, but for the fit on the {\tt cutep50} model of XSPEC.}
\label{batse_sw_sim_apha}
\end{figure}

The amount of GRB spectra with the photon indices harder than $2/3$ depends not only on the  assumed initial distribution of the photon indices, but also on the size of the errors of the reconstructed photon index. The problem is that the statistical scatter of the measurement of the photon index can result in shifts of the values of reconstructed indices toward harder or softer values, thus increasing or decreasing the number of spectra with indices harder than $2/3$. In order to account for this effect we have  performed Monte-Carlo simulations, randomly changing the values of the reconstructed photon indices within the errors and assuming Gaussian probability distribution. We have generated histograms of distributions of the photon indices of the "randomized" datasets, similar to the one shown in Fig. \ref{batse_sw_sim_apha} for the case of the initial simulated BATSE dataset. This has enabled us to estimate the typical statistical scatter of the number of GRB spectra in each bin of the histogram. The statistical errors of the numbers of GRBs with given photon indices, computed in this way, are shown in Fig. \ref{batse_sw_sim_apha}. The uncertainty of the number of GRB spectra with photon indices harder than $2/3$ is given also in the 4th column of the Table \ref{tnumbbeyond}.
 
As one can see from Table \ref{tnumbbeyond}, the limited energy range of \swift\ makes the straightforward confirmation/disproval of BATSE result impossible. Even if one would be able to collect a sample of the GRB spectra comparable to the one of the BATSE spectral database, the large uncertainties in the measurement of the photon index within the {\tt cutep50} model would not allow to claim the presence/absence of the GRB spectra with indices harder than $2/3$ at more than $3\sigma$ level (see 5-th column of Table \ref{tnumbbeyond}). A better constraint would be, in principle, possible if the spectra are fit on the {\tt power50} model, in which the uncertainty of the measurement of the photon index is somewhat lower. In this case, a clear confirmation/disproval of BATSE result at $6\sigma$ level would be possible with a sample of the size comparable to the one of the BATSE spectral database.

However the sample of \swift\ GRBs chosen using the selection criteria similar to the ones used in the BATSE database results in a sample which is $\sim 20$ times smaller than the complete BATSE sample. Indeed, BATSE was an all-sky monitor, while the (partially + fully coded) field of view of BAT is limited to $1.4$~sr \citep{swiftbat}, which is  $4\pi/1.4\simeq 10$ times smaller than that of BATSE. Besides, BATSE telescope operated over the $10$~yr time period, while the BAT has only been 4 years in orbit.  Adopting {\it exactly} identical selection criteria for \swift\ GRBs would, therefore result in selection of  $8\times 10^3\cdot 0.1\cdot (4/10)\simeq 3\times 10^2$ time intervals.  Re-scaling our simulation results to a smaller sample of GRB spectra (see rows 3 and 4 in the Table \ref{tnumbbeyond}, one finds that the number of GRB spectra with reconstructed photon indices harder than $2/3$ is dominated by the statistical uncertainty, so that no definite conclusion can be drawn about presence/absence of such hard spectra, based on the analysis of the histogram of distribution of the photon indices.

\section{Method of testing the BATSE result}
\label{simul1}
 
The large statistical uncertainty of the number of reconstructed GRB spectra with photon indices harder than $2/3$ leaves only a limited possibility to test  the presence of the very hard spectra during the prompt emission phase with \swift. This possibility is related to the fact that in the BATSE GRB spectra sample there exists a certain amount of bursts for which the measured spectra are characterized by low-energy photon indices harder than $2/3$ at $\ge 3\sigma$ confidence level.  One expects to find similar high-confidence hard spectra in a sub-set of the time resolved \swift\ GRB spectra. Comparing the number of detected hard GRB spectra with the one expected from the simulations of the BATSE GRB spectral sample, one can readily test BATSE result.  Contrary to the method based on the analysis of the distribution of the low-energy photon indices, discussed above, such a "direct" method does not suffer from the statistical uncertainties of the distribution of the photon indices. 

A potential problem of the proposed method is that the episodes of extreme hardening of the spectra are not randomly distributed over the BATSE GRB sample. In fact, the episodes of hardening last for more than 1 time bin, so that individual GRBs in which the hard indices are found, posses multiple time intervals with hard spectra. This is clear from the last column in Table \ref{tnumbbeyond}. One can see that in spite of the considerable amount of time intervals with hard spectra (28 in the case of {\tt cutep50} model) expected in \swift\ GRB sample, these intervals are expected to be found in only $\sim 5$~GRBs. This means that the GRBs with hard photon indices are extremely rare events for \swift. One expects to find $\sim 1$ such event per year. The expectation is still worse, if one adopts the {\tt power50} model for the spectral analysis. In this case one expects to detect one such event per 2 years on average.  In the following section we report the results of a systematic search of the  episodes of hardening of spectra beyond $\Gamma=2/3$ in the real data of BAT telescope on board of \swift.

\section{Selection of \swift\ GRBs for the time-resolved spectral analysis}
\label{sec:selection}

The BATSE  sample of the time resolved GRB spectra which we have used for our simulations in the previous section was selected from sufficiently bright bursts detected over a 10~year span of the {\it CGRO} mission according to the selection criteria  \citep{kaneko08}:
\begin{itemize}
\item (A) energy fluence in $20-2000$~keV energy band is larger than $2\times 10^{-5}$~erg/cm$^2$s or
\item (B) peak flux in a 256~ms time bin is higher than $10$~ph/cm$^2$s in $50-300$~keV energy band.
\end{itemize}
Each of the selected BATSE GRBs was binned in time intervals for the time-resolved spectral analysis, according to the following principle:
\begin{itemize}
\item (C) Signal to noise ratio in the time interval is $\ge 45$.
\end{itemize}

\begin{figure}
\includegraphics[width=1.0\columnwidth,angle=0]{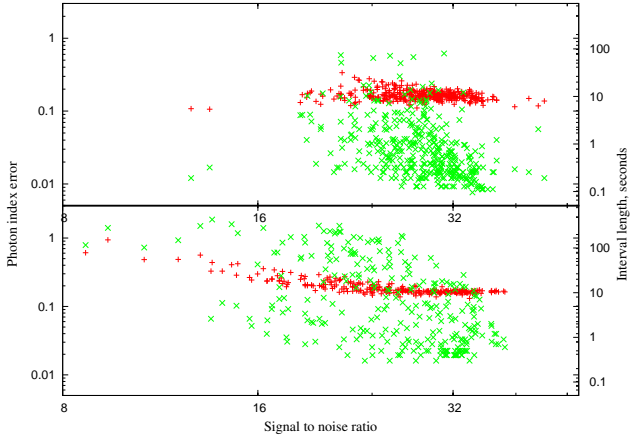}
\caption{Errors of measurement of the photon index (red / dark grey) and the durations of the time intervals selected for the spectral extraction (green / light grey points) as functions of the signal to noise ratios.  The upper panel shows the simulated BATSE spectra. The lower panel shows the real \swift\ data. }
\label{spectra_errors}
\end{figure}

Obviously, because of the difference in the \swift/BAT and BATSE energy bands, it is not possible to adopt literally the same selection criteria (A), (B) for selection of \swift\ GRBs. However,  it is straightforward to work out a selection criteria for the time intervals, equivalent to (C), using the sample of simulated BATSE GRBs. 
The upper panel of Fig. \ref{spectra_errors} shows the distribution of signal to noise ratios $S/N$ (expressed in the "total counts", i.e. $(S/N)^2$ plotted over $x$ axis) for the simulated BATSE GRB spectra, together with the durations of the time intervals and errors of the measurement of the photon indices (adopting model power50). One can see that the time binning which corresponds to $S/N=45$ in BATSE is equivalent to a similar $S/N\sim 30-40$, or to the photon statistics $\sim 10^3$  in \swift. 

To select the time intervals with signal to noise ratio $\sim 30-40$ from the real \swift\ data we adopt the following procedure. The BAT lightcurves are usually presented in terms of the "normalized counts", which are the photon counts which would be detected from an equivalent on-axis source, rather than from the real GRB source at non-zero off-axis angle. Selection of the time intervals for the spectral analysis in terms of the photon statistics estimated from the "normalized counts" usually over-estimates the signal to noise ratio, because (a) the burst could initially occur in the partially coded field of view and (b) the "normalized counts" statistics does not take into account the systematic error of the photon flux measurement. Having this in mind, we have found that re-formulating the time binning principle as  
\begin{itemize}
\item  (C1) Signal to noise ratio in the time interval for spectral extraction is $\ge 60$
\end{itemize}
results in a distribution of the errors of measurement of the photon indices in the real \swift\ data, which is similar to the ones of the simulated BATSE spectra. This is demonstrated in the lower panel of Fig. \ref{spectra_errors} where the signal to noise ratios, errors of the measurement of the powerlaw index and durations of the time intervals selected in the real \swift\ GRB spectra are shown. Comparing the lower and upper panels of Fig. \ref{spectra_errors}, one finds that main difference between the simulated BATSE and real \swift\ spectra, selected according to the above criteria (C) and (C1),  respectively,  is in the duration of the time intervals. 
Somewhat longer durations of the time intervals in the real \swift\ data are explained by the fact that in our simulations of BATSE spectra we have ignored the possibility that the GRB can occur at large off-axis angle and/or because the simulation procedure occasionally under-estimates the systematic measurement errors. 

Imposing a simple selection criterion on the GRBs
\begin{itemize}
\item  (AB1) GRB contains at least one time interval suitable for spectral extraction,
\end{itemize}
we have selected  a set of 80 \swift\ GRBs, listed in Table \ref{tab:grbs}. Binning the selected GRBs onto time intervals for the spectral extraction we have obtained $\sim 3\times 10^2$ time resolved spectra (see Table \ref{tnumbbeyond}). As expected, the overall size of the \swift\ GRB spectral sample is some $\sim 20$ times smaller than the one of the BATSE spectral database.

The selection criterion (AB1) adopted for \swift\ GRBs is weaker than the  criteria (A,B) used for the selection of BATSE GRBs. This means that, in principle, one should select more GRBs for the spectral analysis, than expected from the simple re-scaling of the BATSE GRB sample. An over-estimate of the expected number of \swift\ GRBs satisfying the selection criteria of BATSE GRBs, found in the previous section, is related to the fact that  we have ignored the decrease of the telescope sensitivity in the partially coded field of view. However, based on the estimates of the previous section, one can see that already with the relaxed selection criterion (AB1), which reduces the amount of the time intervals available for the spectral analysis down to $\sim 300$, one is left with a very limited possibility to test the BATSE result on the existence of very hard GRB spectra. Taking this into account, we limit our selection criteria to (AB1) and adopt the requirement (C1) for the choice of time binning of the selected GRBs.

\begin{figure}
\includegraphics[width=\columnwidth]{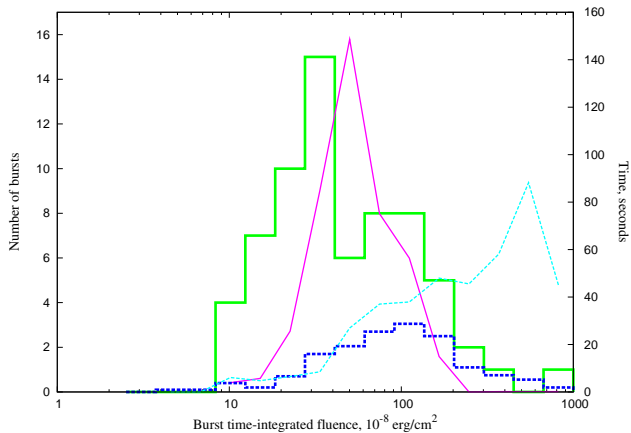}
\caption{Distribution of the GRB fluences. Green thick solid line: Swift GRB fluences; Blue dashed thick line: BATSE GRBs in Swift BAT (number of events rescaled to Swift effective exposure). Thin solid line - median duration of the bursts included in current bin in the \swift\ data, thin dashed line: durations of the bursts in the simulated BATSE GRB sample.}
\label{fluences_hist}
\end{figure}

Relaxing the criteria for the GRB selection, in fact, changes the GRB sample under consideration. To illustrate this fact, we plot in Figure   \ref{fluences_hist} the distribution of the overall fluences of the selected \swift\ GRBs, as compared to the one of the BATSE GRBs used for the spectral analysis by \citet{kaneko08}. One can see that relaxing the selection criteria we add a large amount of bursts with low fluences. Another important difference is that the time binning of the additional lower fluence GRBs is  characterized by longer typical integration times of the spectra (shown by the magenta dotted line in Fig. \ref{fluences_hist}).  The presence of an additional population of weaker GRBs in the sample is not a problem for the method described in Section \ref{simul1}: one looks for episodes of high confidence level spectral hardening in several individual bursts rather than studies statistical properties of the entire selected GRB sample.

\begin{table*}
\label{tab:grbs}
{\footnotesize
\begin{tabular}{lcc|lcc|lcc|lcc}
\hline
Name & Fluence & $N$&Name & Fluence & $N$  &Name &  Fluence & $N$  & Name & Fluence & $N$  \\
\hline
GRB041223 & 163 & 7  & GRB041224 & 54.9 & 1  & GRB050117 & 65.3 & 2  & GRB050219B & 95.3 & 1  \\ 
GRB050306 & 111 & 1  & GRB050418 & 41 & 1  & GRB050525A & 152 & 20  & GRB050713A & 28.3 & 1  \\ 
GRB050717 & 53.5 & 2  & GRB050724 & 11.2 & 1  & GRB050820B & 16.1 & 2  & GRB050904 & 45.3 & 1  \\ 
GRB050915B & 34 & 2  & GRB050922C & 15.7 & 1  & GRB051111 & 25.6 & 1  & GRB051117A & 31.6 & 1  \\ 
GRB051221A & 9.29 & 1  & GRB060105 & 134 & 5  & GRB060117 & 161 & 7  & GRB060210 & 60 & 1  \\ 
GRB060322 & 33.8 & 1  & GRB060413 & 21.1 & 1  & GRB060418 & 67.7 & 2  & GRB060510B & 27.9 & 1  \\ 
GRB060607A & 22.6 & 1  & GRB060614 & 198 & 16  & GRB060729 & 27 & 1  & GRB060813 & 16.9 & 1  \\ 
GRB060929 & 5.31 & 1  & GRB061007 & 425 & 31 & GRB061021 & 14.6 & 1  & GRB061121 & 125 & 17  \\ 
GRB061126 & 24.5 & 1  & GRB061202 & 24.6 & 2  & GRB061222A & 68.4 & 6  & GRB070107 & 41.4 & 1  \\ 
GRB070129 & 20.4 & 1  & GRB070220 & 86.1 & 4  & GRB070306 & 28 & 2  & GRB070328 & 17.9 & 1  \\ 
GRB070419B & 69.3 & 3  & GRB070420 & 118 & 1  & GRB070521 & 76.9 & 4  & GRB070612A & 97.5 & 1  \\ 
GRB070616 & 181 & 8  & GRB070621 & 44.9 & 1  & GRB070721B & 28.5 & 1  & GRB070911 & 104 & 7  \\ 
GRB070917 & 9.82 & 1  & GRB071003 & 35.5 & 1  & GRB071010B & 19.7 & 3  & GRB071020 & 17.3 & 2  \\ 
GRB071117 & 15 & 1  & GRB080207 & 30.4 & 1  & GRB080229A & 71 & 3  & GRB080310 & 18.6 & 1  \\ 
GRB080319B & 772 & 96  & GRB080328 & 79.5 & 3  & GRB080411 & 253 & 31  & GRB080413A & 32 & 2  \\ 
GRB080413B & 25.9 & 1  & GRB080503 & 19.3 & 1  & GRB080603B & 12 & 1  & GRB080605 & 123 & 6  \\ 
GRB080607 & 222 & 4  & GRB080613B & 37.5 & 2  & GRB080810 & 33.1 & 1  & GRB080916A & 33.1 & 2  \\ 
GRB080928 & 18.4 & 1  & GRB081008 & 29.6 & 1  & GRB081028A & 32.2 & 1  & & & \\

\hline
\end{tabular}
} \caption{Swift gamma-ray bursts selected for the analysis.}
\end{table*}

\section{Data reduction and analysis }\label{method}

To analyze the set of selected \swift\ GRBs in a homogeneous way, we have developed a pipeline script (which is similar to {\it batgrbproduct} pipeline script provided by the \swift\ team). We use $heasoft-6.5.1$ and latest calibration database in our analysis. 

Using BAT science analysis tasks {\it batmaskwtevt} (applying detector mask corrections from auxiliary files and estimated by detector plain image hot pixels) and {\it batbinevt} we produce  "mask tagged" lightcurves for
each burst selected and for different time binning (we select time
binning intervals as $5,100,500$ and $2000$~ msec for each
burst). Auxiliary \swift\ slew information extracted during mask weighting is saved for further use.
 The lightcurves are built in units of background subtracted counts per fully illuminated detector for an equivalent on-axis source, or "normalized counts", as it is defined in \swift\ User Guide.
Overall time intervals of the burst lightcurves considered in our analysis  were selected with the help of {\it battblocks} tool \textsl{T99} (time interval containing $99\%$ of GRB fluence). We choose knowingly too long time around real GRB and reject remaining background intervals (identified as intervals with low peak flux) after analysis.

At the next stage of analysis, the GRB lightcurves are re-binned to achieve the required signal to noise ratio, given by the selection criterion (C1) (see section \ref{simul1}).  The significance of the signal detection normally can not be extracted directly from the information about the "normalized counts" found in the lightcurves. This is related to the fact that the flux measurement error can be determined only after background subtraction via mask weighting technique and taking into account the systematic errors.  In order to characterize the significance of the signal detection, we introduce the "real counts" in each time bin, via the relation $C_{real}=\left(F/\Delta F\right)^2$, where $F$ is the flux and $\Delta F$ is the flux measurement error. Each time bin selected for the spectral analysis is required to accumulate the number of real counts above a certain threshold. We perform the analysis via several parallel pipelines, which differ by the choice of the required number of real counts per time bin. Choosing the binning with a relatively small number of real counts, 100, 400 and 900  per time bin, we are able to trace the spectral evolution of the bursts on shorter time scales. On the other side, choosing a large number of real counts per time bin enables to obtain higher quality spectra. 

The choice of 4000 real counts per bin (criterion C1 for the GRB time binning) results in the quality of reconstruction of spectral parameters from the BAT spectra, which is comparable to the one found in the simulated BATSE spectra. This is illustrated in the lower panel of Fig. \ref{spectra_errors}, where the errors of measurement of the photon index as a function of significance of detection is shown. One could notice that the significance of the signal detection found from the BAT spectrum (plotted along the $x$ axis in Fig. \ref{spectra_errors} is systematically lower than the one found from the BAT lightcurve (assumed to be constant in all time bins, $\sim \sqrt{4000}\simeq 63$). This is, related to the fact that the \swift\ spectral and lightcurve extraction procedures use different estimates of systematic errors. 

Having binned the GRB in time intervals with sufficient signal to noise ratio, we perform spectral extraction procedure in each time bin. The entire 15-150~keV energy interval is divided onto 10
energy bins: 15-25~keV, 25-35~keV, 35-45~keV, 45-55~keV, 55-65~keV, 65-75~keV, 75-90~keV, 90-105~keV, 105-125~keV, 125-150~keV. The spectra are extracted with the help of the {\it batbinevt} tool.  The response matrices are produced with the help of {\it batdrmgen} tool.  We insert the proper
ray-tracing keywords into the DRM file with the help of {\it batupdatephakw}. The systematic error vector is produced with the help of {\it batphasyerr} and  applied to the
spectrum\footnote{see for details http://swift.gsfc.nasa.gov/docs/swift/analysis/bat\_digest.html}.

The individual spectra are analyzed with the help of XSPEC  package\footnote{XSPEC reference}. The spectral fits are done using both {\tt power50} (powerlaw normalized at 50~keV) model and cut-off power law {\tt cutep50} (powerlaw with high-energy cutoff, normalized at 50~keV) model of XSPEC, recommended by the BAT team\footnote{http://heasarc.gsfc.nasa.gov/docs/swift/analysis/threads/batspectrumthread.html}. As discussed above, in the energy range covered by \swift, the phenomenological Band model for the spectrum (\ref{band}) is equivalent to the cut-off power-law \texttt{cutep50}, with the photon index and cut-off energy simply related to the low-energy photon index $\alpha$ and the break energy $E_{\rm break}$ of the Band model, see Eq. (\ref{alpha_Ecut}). 

The cutoff energy $E_{\rm cut}$ is higher then $150$~keV for a large fraction of GRBs. 
This means that this energy can not be fully constrained by \swift/BAT. Taking this into account, we fit the spectra with 2 alternative models: {\tt cutep50} and {\tt power50}, the latter being appropriate when $E_{\rm cut}$ lies much above the BAT energy band. 

\begin{figure}
\includegraphics[width=0.8\columnwidth,angle=-90]{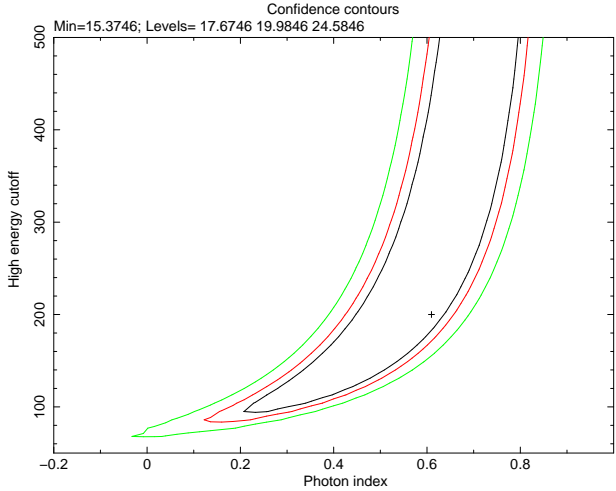}
\caption{Chi-squared levels for fitting a typical BAT spectrum with cutoff powerlaw model.
\textsl{Swift} GRB spectra, fitted on \textsl{cutep50} model.}
\label{ecut_contour}
\end{figure}

It is clear that fitting the spectra with a cutoff powerlaw model results in a larger uncertainty of the measurement of the photon index, because of a parameter degeneracy arising in simultaneous measurement of the photon spectra index and the cutoff energy. This effect is illustrated in Fig. \ref{ecut_contour} where we show 1 2 and 3 sigma error contours in the $E_{\rm cut}, \Gamma$ plane for a typical GRB spectrum. One can see that although the cut-off energy is not constrained by the fit, it affects the error of the photon index, effectively increasing it by a factor of 2. The uncertainty of calculation of the photon index is much smaller in the fits on the powerlaw model, where the parameter $E_{\rm cut}$ is absent.

\section{ \swift\ GRBs beyond the "synchrotron deathline"}\label{res}

\subsection{ Spectral evolution of individual \swift\ bursts }

The entire set of \swift\ GRBs selected according to the criterion (AB1) and time binned following the criterion (C1) (see Section \ref{simulations}) reveals only two GRBs in which in total 5 time intervals  with hardening of the spectrum down to the photon indices $\Gamma<2/3$ at $\ge 3\sigma$ level is observed. This are  GRB061007 and GRB 080319B. The lightcurves of these GRBs and the evolution of the photon index over the burst duration are shown in Fig. \ref{lightcurves}. 

\subsection{GRB 080319B}

is the GRB with the highest fluence in the \swift\ GRB sample  (see Table \ref{tab:grbs}). The period of spectral hardening beyond $\Gamma=2/3$ occurs during the first  10~s of the burst duration (see Fig. \ref{lightcurves}a). GRB 080319B is one of the few GRBs for which a bright prompt optical emission was detected \citep{racusin08}. The optical emission flux is above the powerlaw extrapolation of the observed \gr\ flux. This indicates that the optical emission forms a separate component which is either produced by a separate population of relativistic electrons or, otherwise, it can be produced by the same population of electrons, but via a different physical mechanism. It is interesting to note that the moment of on-set of the flash coincides with the moment of softening of the spectrum from $\Gamma\le 2/3$ down to $\Gamma\simeq 1$ (see Fig. \ref{lightcurves}a). This gives a clue to the understanding of the mechanism of the strong optical emission \citep{savchenko08} and, possibly, for the understanding of the mechanism of production of very hard \gr\ emission spectrum during the first 10~s of the burst.

\subsection{GRB 061007}

is the GRB with the second highest fluence in the \swift\ GRB sample. It is remarkable that the episodes of spectral hardening down to the photon index $\Gamma\le 2/3$ are detected in the two brightest GRBs. This is consistent with our conclusion that the sensitivity of BAT telescope is only marginally sufficient for the detection of the very hard GRB spectra: the study of the spectral evolution with sufficient time resolution is possible only for the brightest bursts. Contrary to GRB 080319B, the episode of hardening of the spectrum down to $\Gamma\le 2/3$ in GRB 061007 is observed in the middle, rather than at the beginning of the burst. However, one could note that the episode of the hardening coincides with the start of a pronounced sub-flare (see Fig. \ref{lightcurves}b). 

\subsection{Other GRBs with the episodes of spectral hardening}

The fact that only two brightest bursts from our \swift\ GRB sample reveal the episodes of spectral hardening beyond $\Gamma=2/3$ may indicate that our requirement (C1) adopted for the time binning of the GRB lightcurve is too restrictive. The typical width of the time bins chosen for the spectral analysis in weaker GRBs turns out to be longer than the typical duration of the episodes of the spectral hardening. It is possible that relaxing the constraint on the overall significance of the signal detection within the time bin one may be able to find more episodes of spectral hardening in the time resolved GRB spectra. 

\begin{figure*}
\includegraphics[width=0.65\columnwidth,angle=0]{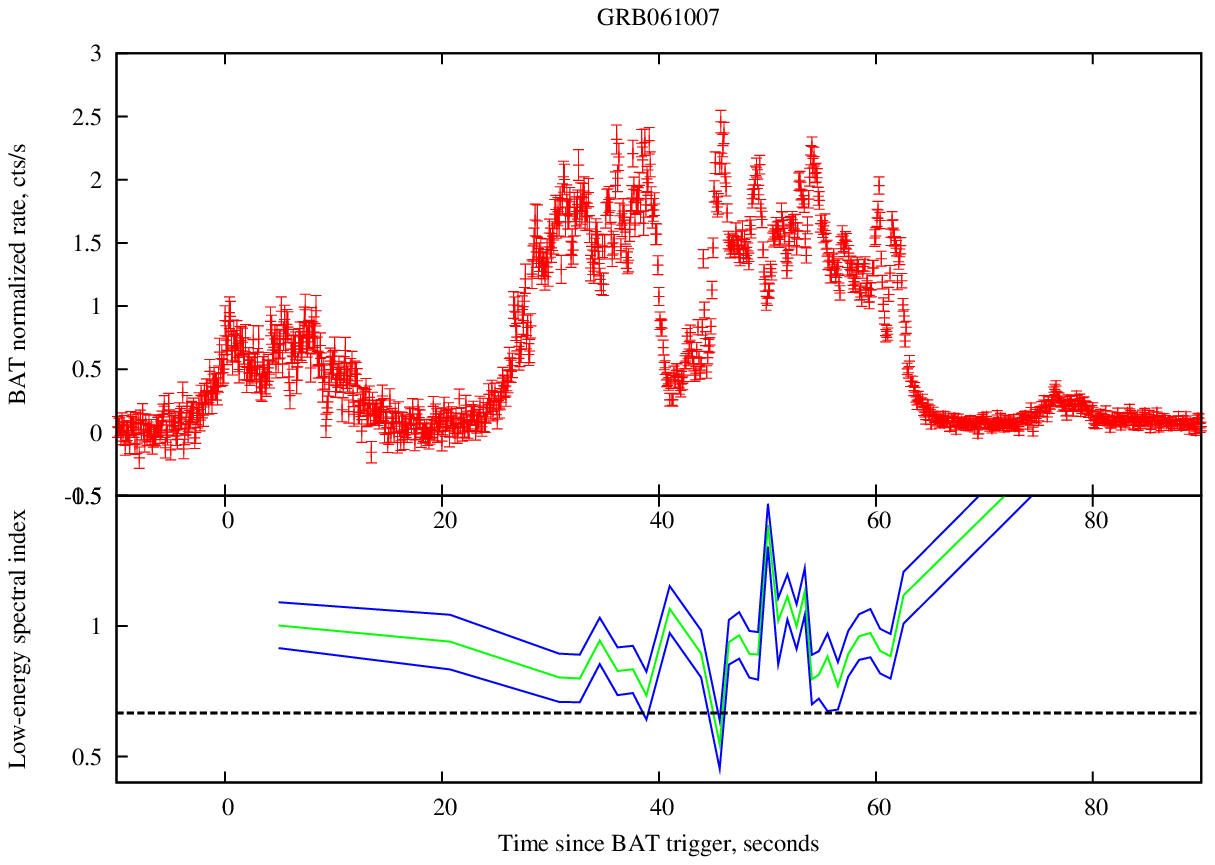}
\includegraphics[width=0.65\columnwidth,angle=0]{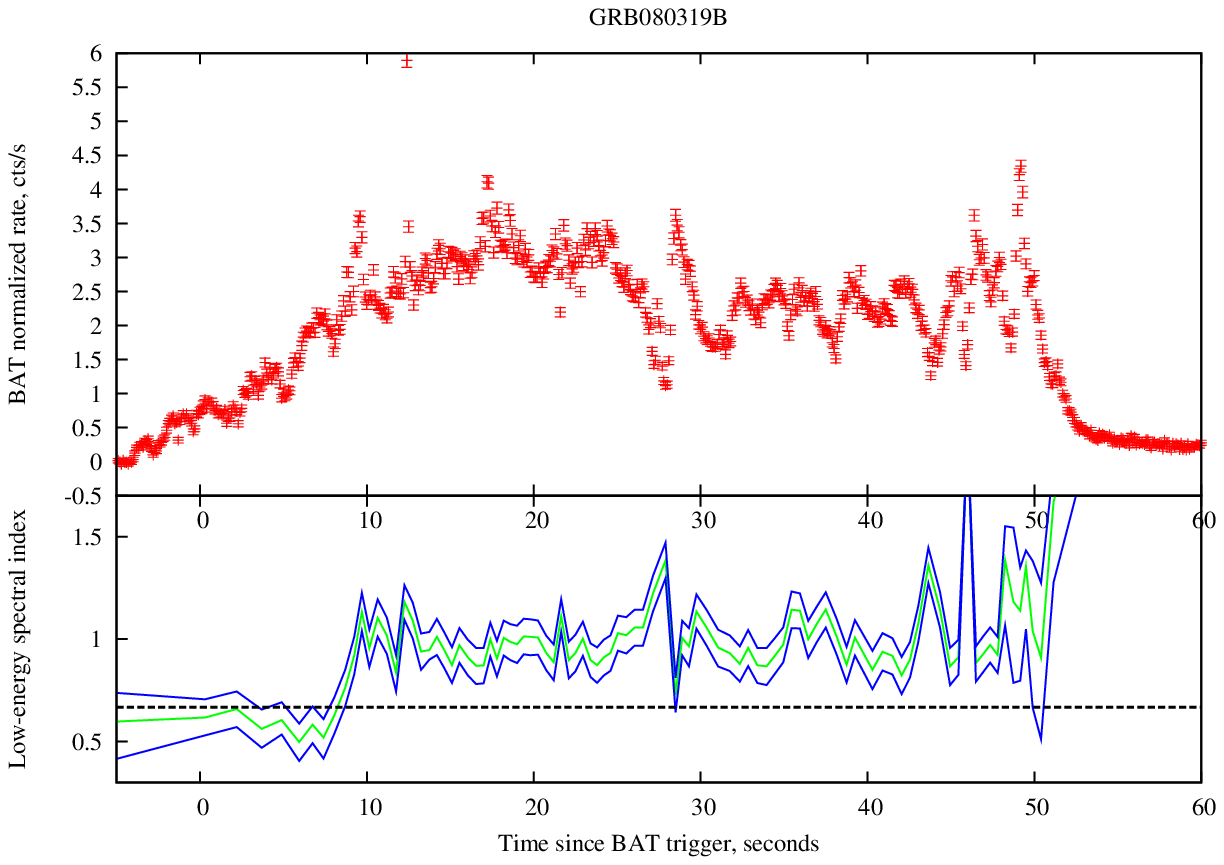}
\includegraphics[width=0.65\columnwidth,angle=0]{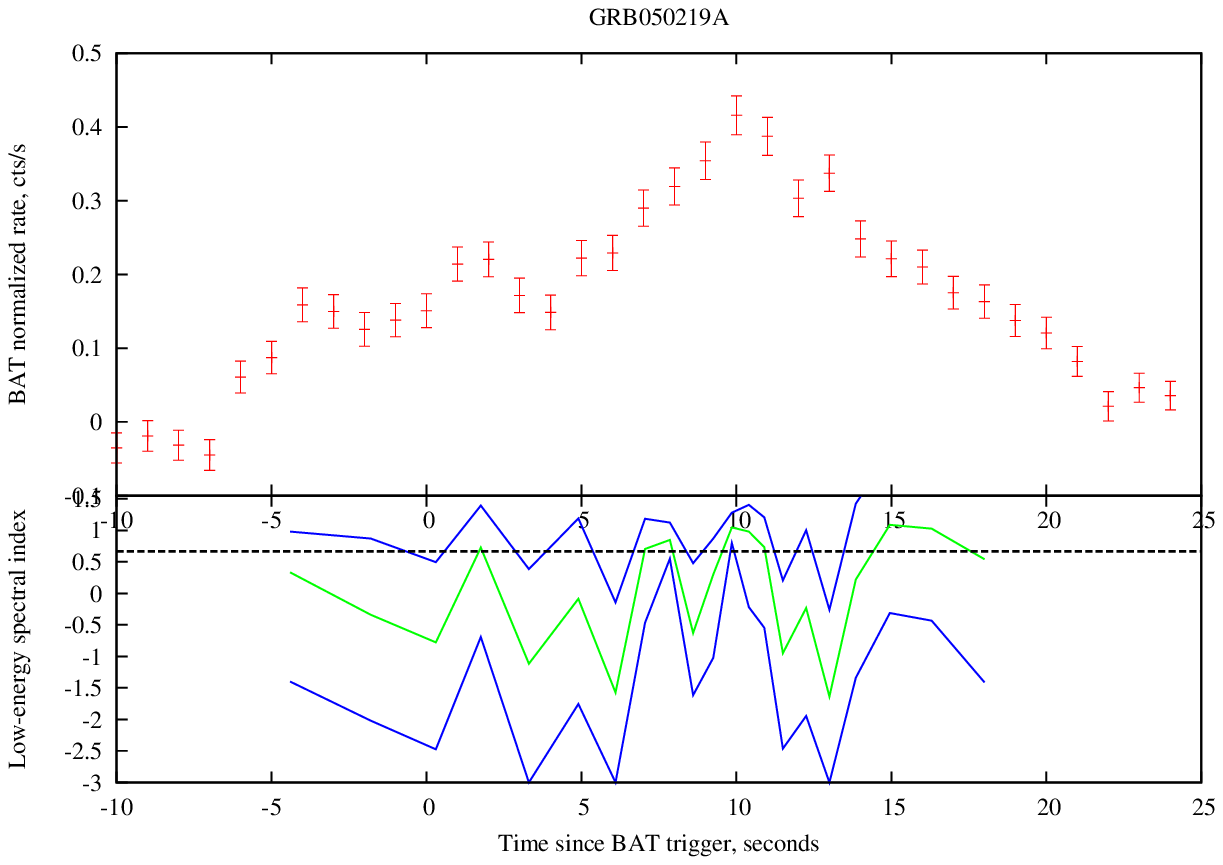}
\includegraphics[width=0.65\columnwidth,angle=0]{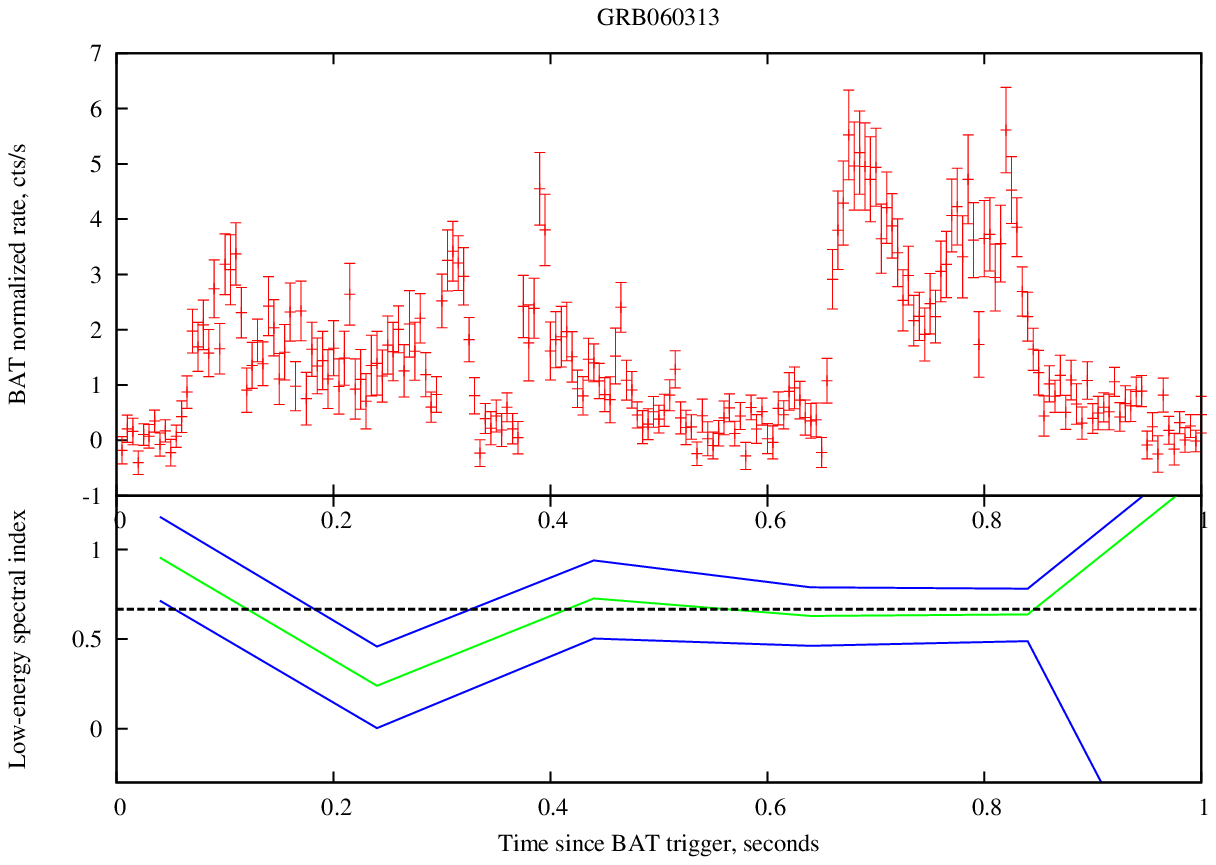}
\includegraphics[width=0.65\columnwidth,angle=0]{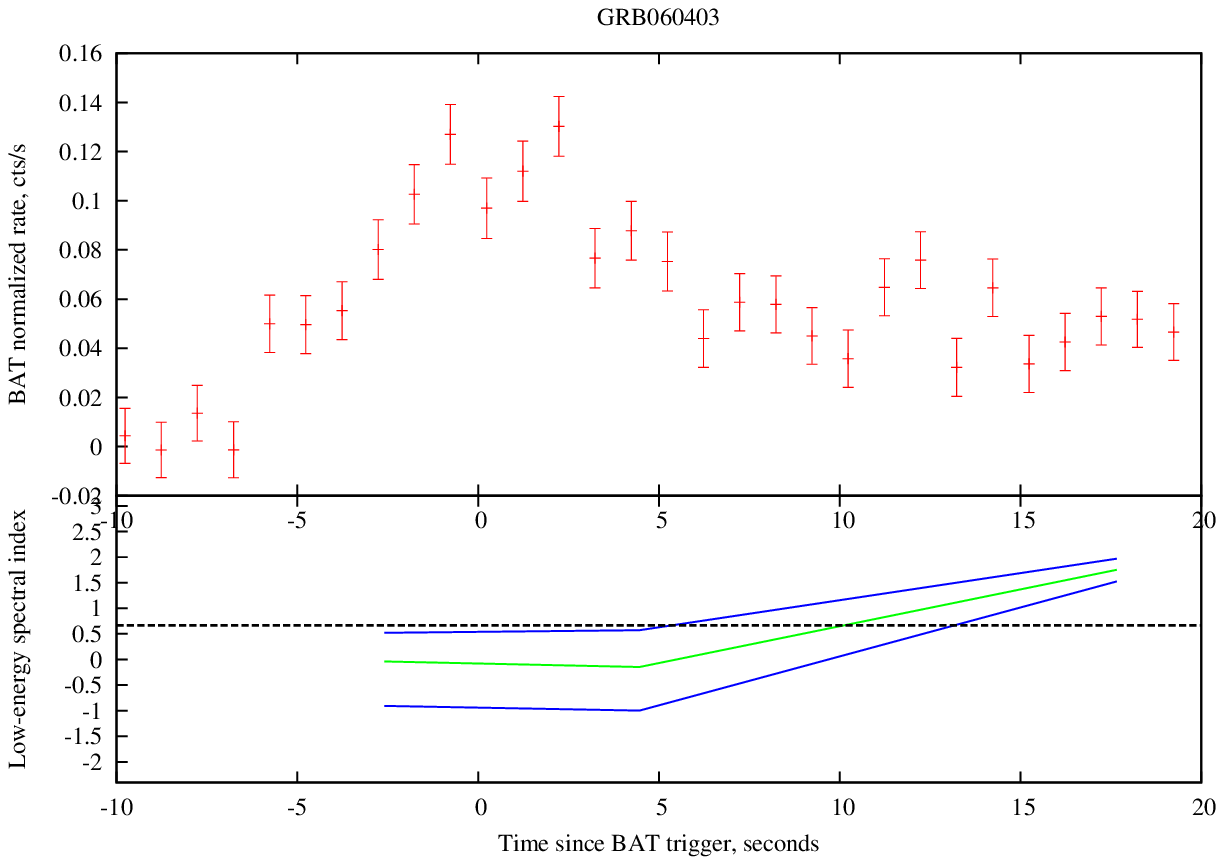}
\includegraphics[width=0.65\columnwidth,angle=0]{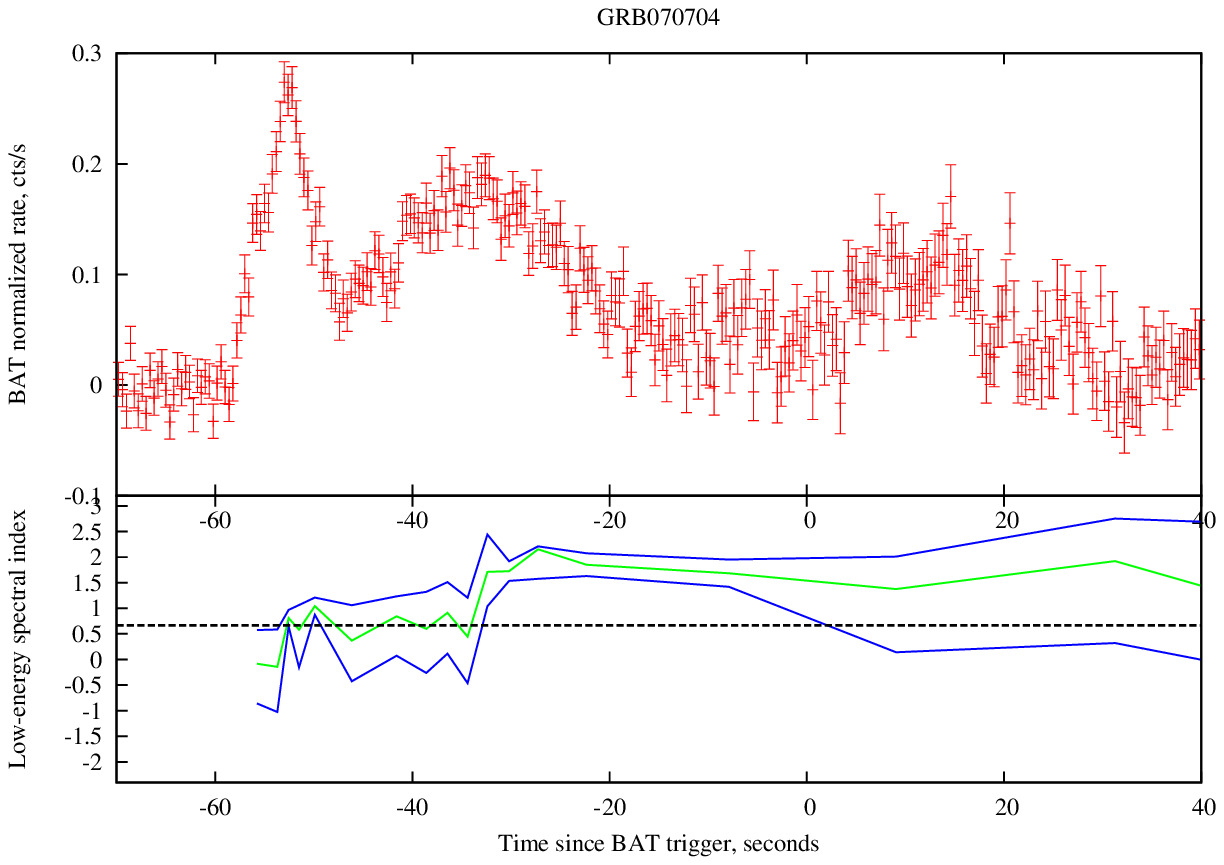}
\includegraphics[width=0.65\columnwidth,angle=0]{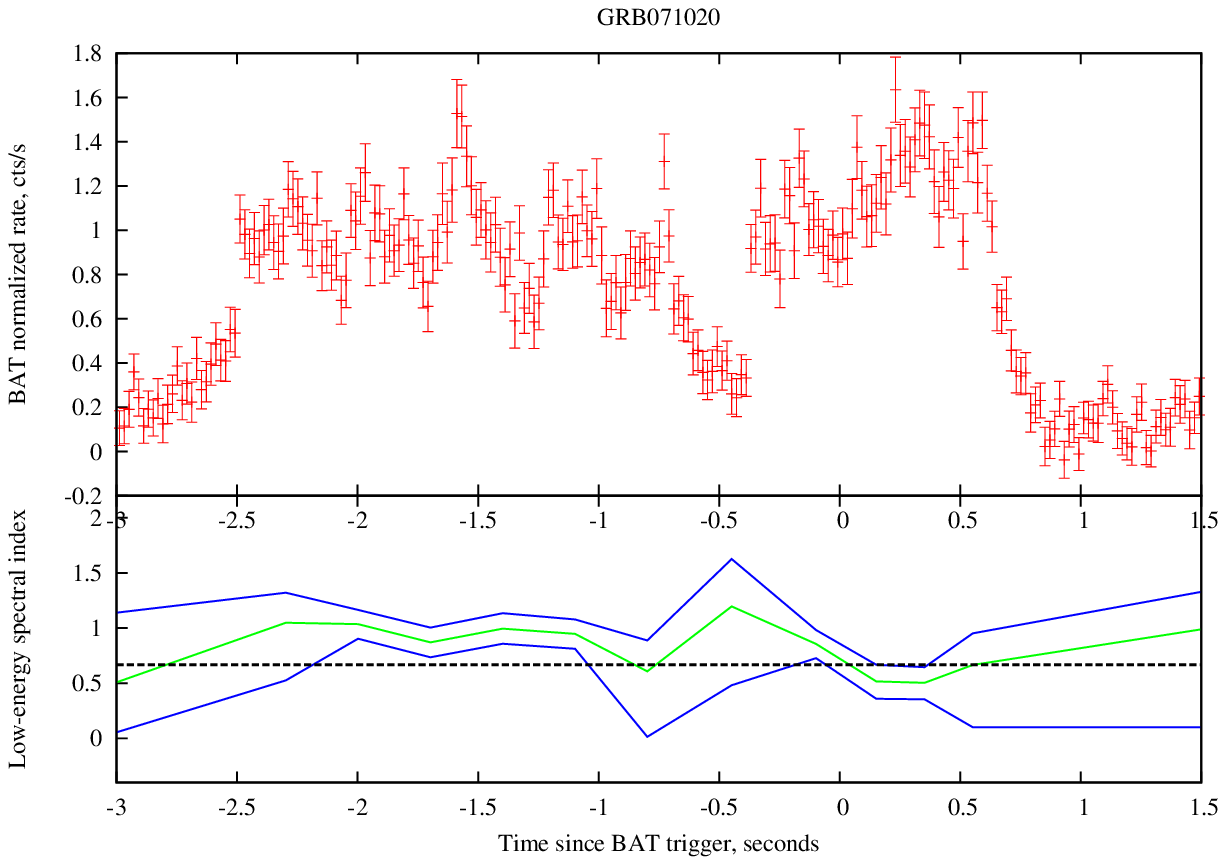}
\includegraphics[width=0.65\columnwidth,angle=0]{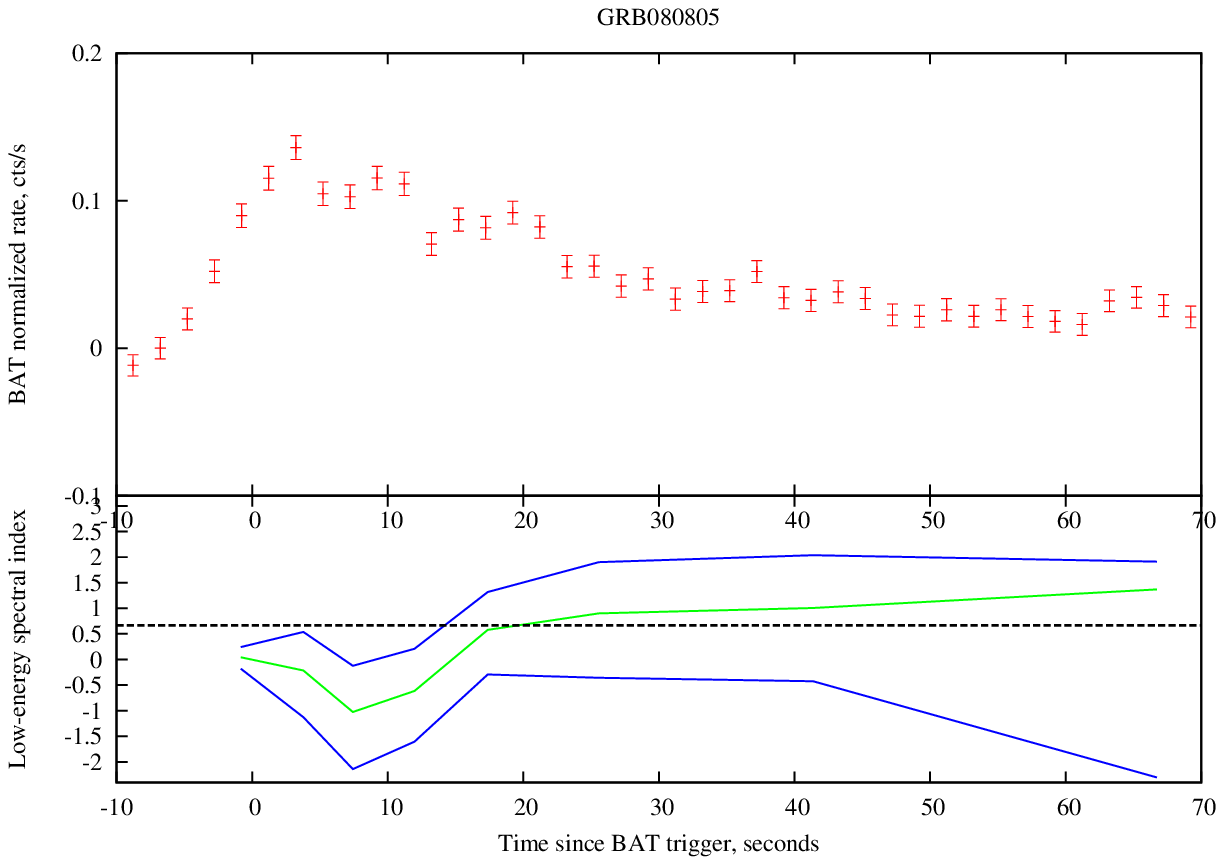}
\includegraphics[width=0.65\columnwidth,angle=0]{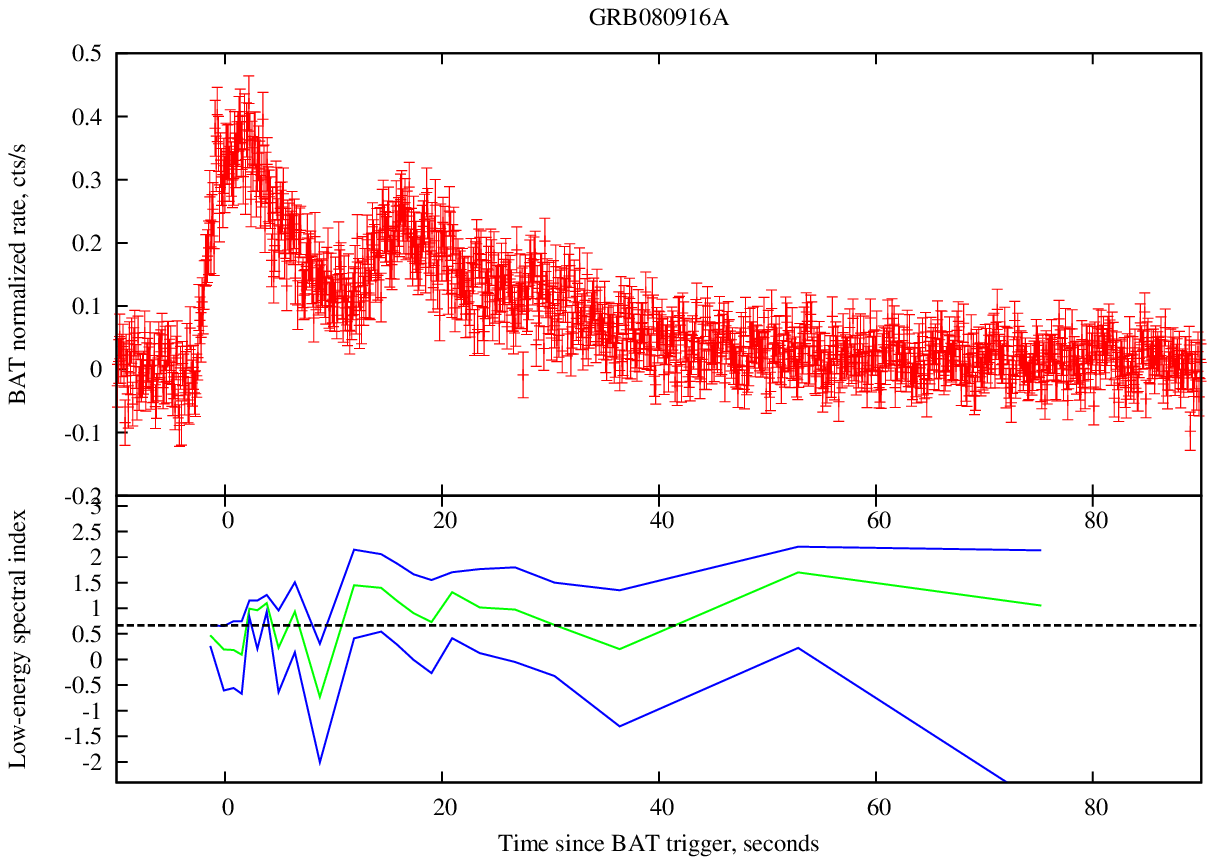}
\caption{Episodes of spectral hardening beyond $\Gamma=2/3$ in \swift\ GRBs. Top panels show the 15-150~keV band lightcurves. Bottom panels show the evolution of the photon index in the {\tt cutep50} model (green curves). Blue curves show the $3\sigma$ errors of the measurement of the photon index.
Horizontal dashed line in the bottom panels shows the limiting photon index of the optically thin synchrotron emission, $\Gamma=2/3$.}
\label{lightcurves}
\end{figure*}

This expectation is confirmed by our re analysis of the data with different choice of the time binning, which reveals other GRBs with the episodes of spectral hardening, GRB 050219A, GRB 060313, GRB 060403, GRB 070704, GRB 071020, GRB 080805 and GRB 080916A, shown in panels c-i of Fig. \ref{lightcurves}.

To the best of our knowledge, not all the above listed episodes of spectral hardening down to $\Gamma<2/3$ were previously reported. In the case of GRB 050219A the spectral hardening is discussed by \citet{goad_05}.  GRB 060313 is listed as hardest of the \swift\ bursts (based on the $(50-100 keV)/(25- 50 keV)$ ratio) in 
\citep{roming_06}).  

\begin{figure}
\includegraphics[width=1.05\columnwidth,angle=0]{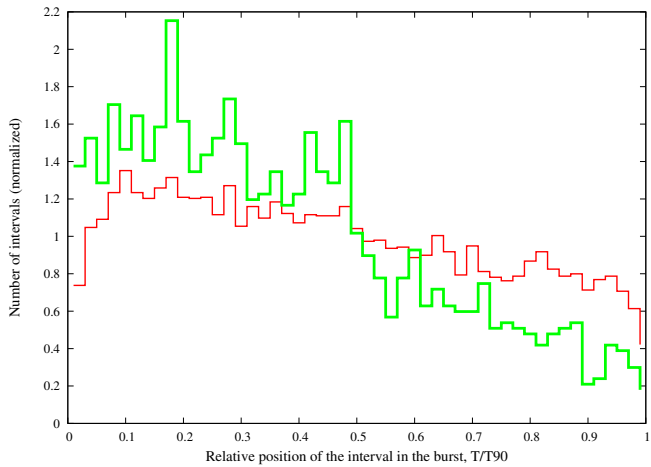}
\caption{Distribution of the times episodes of hardening of the spectra of BATSE GRBs beyond $\Gamma=2/3$ over the duration of the GRB (green) compared to the overall distribution of the numbers of time resolved GRB spectra throughout the GRB duration (red).  }
\label{batse_htimes}
\end{figure}

The episodes of spectral hardening do appear preferentially at the on-set of the bursts or of the bright sub-flares within the bursts. The same effect is observed in the case of BATSE GRBs, where the episodes of harder than $\Gamma=2/3$ spectra appear preferentially at the beginning of the burst (see Fig. \ref{batse_htimes}). 

\begin{figure}
\includegraphics[width=1.05\columnwidth,angle=0]{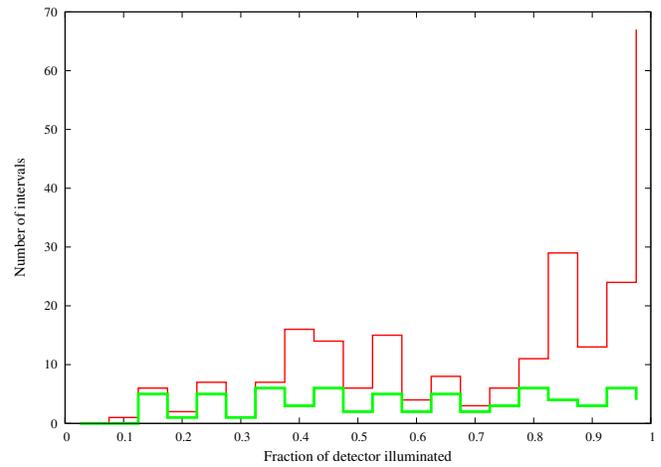}
\caption{Distribution of the illumination fractions of BAT detector surface .}
\label{pcode}
\end{figure}

The appearance of the episodes of spectral hardening preferentially at the on-set of the bursts reveals an additional difficulty for detection of such intervals with \swift. Namely,  a significant fraction of the GRBs is initially detected in the partially coded field of view of the BAT telescope (see Fig \ref{pcode}). In this part of the field of view the sensitivity of the telescope degrades, which leads to a decrease of the signal statistics. This results in a decrease of the precision of the measurement of spectral parameters. This, in turn, leads to the decrease of efficiency of detection of the episodes of spectral hardening at the on-set of the burst. This effect was not included in our simulations of BATSE GRB spectra. As a result, our prediction of the expected number of GRB spectra harder than $\Gamma=2/3$ at $\ge 3\sigma$ confidence level, based on the simulations of BATSE spectra, is, in fact, an over-estimate.

To find out how the under-estimate of the error of measurement of the photon index in the simulations of BATSE spectra influences the prediction of the amount of GRB spectra beyond the "synchrotron deathline", we give in Table \ref{tnumbbeyond} (simulated GRB sample called "BATSE, scaled (2)") the results of simulation of BATSE GRB spectra with BAT response, assuming that the error of measurement of the photon index is 2 times larger than the one predicted by the our initial simulation algorithm, described in Section \ref{simulations}.  One can see that a factor of 2 increase of the measurement error results in a dramatic decrease of the expected detections of the episodes of spectral hardening ($\sim 2$ time intervals in $\sim 1$ GRB in 4 years of \swift\ data). 

Taking into account this uncertainty on the error of the measurement of the photon indices in the simulations of BATSE GRBs, one can conclude that the detection of 5 intervals of hardening of the spectra beyond $\Gamma=2/3$ at $\ge 3\sigma$ confidence level in \swift\ GRB dataset is consistent with the expectations derived from the simulations of observational appearance of BATSE GRBs in BAT. 

\subsection{Distribution of low-energy photon indices in \swift\ GRBs}

In this sub-section we present the distribution of the GRB photon indices derived from the real \swift\ data for the two spectral models ({\tt power50} and {\tt cutep50}) used in our spectral analysis. As it is mentioned above, we have relaxed the GRB selection criteria for \swift\ GRBs to the single criterion (AB1), compared to the tighter criteria (A),(B) used for the BATSE GRBs in the sample studied by \citet{kaneko08}. This has resulted in the inclusion of lower fluence and lower flux GRBs in our sample (see Fig. \ref{fluences_hist}). 

\begin{figure}
\includegraphics[width=1.00\columnwidth,angle=0]{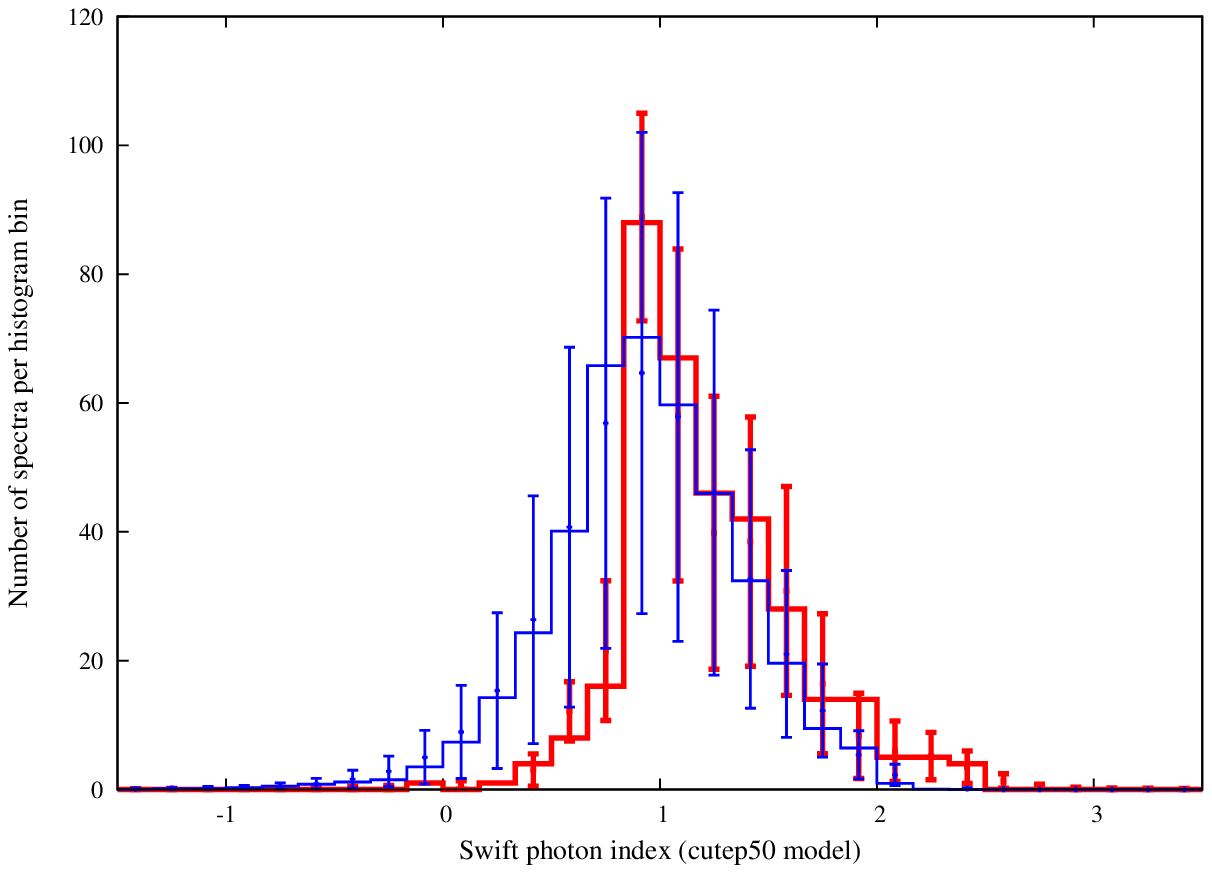}
\includegraphics[width=1.\columnwidth,angle=0]{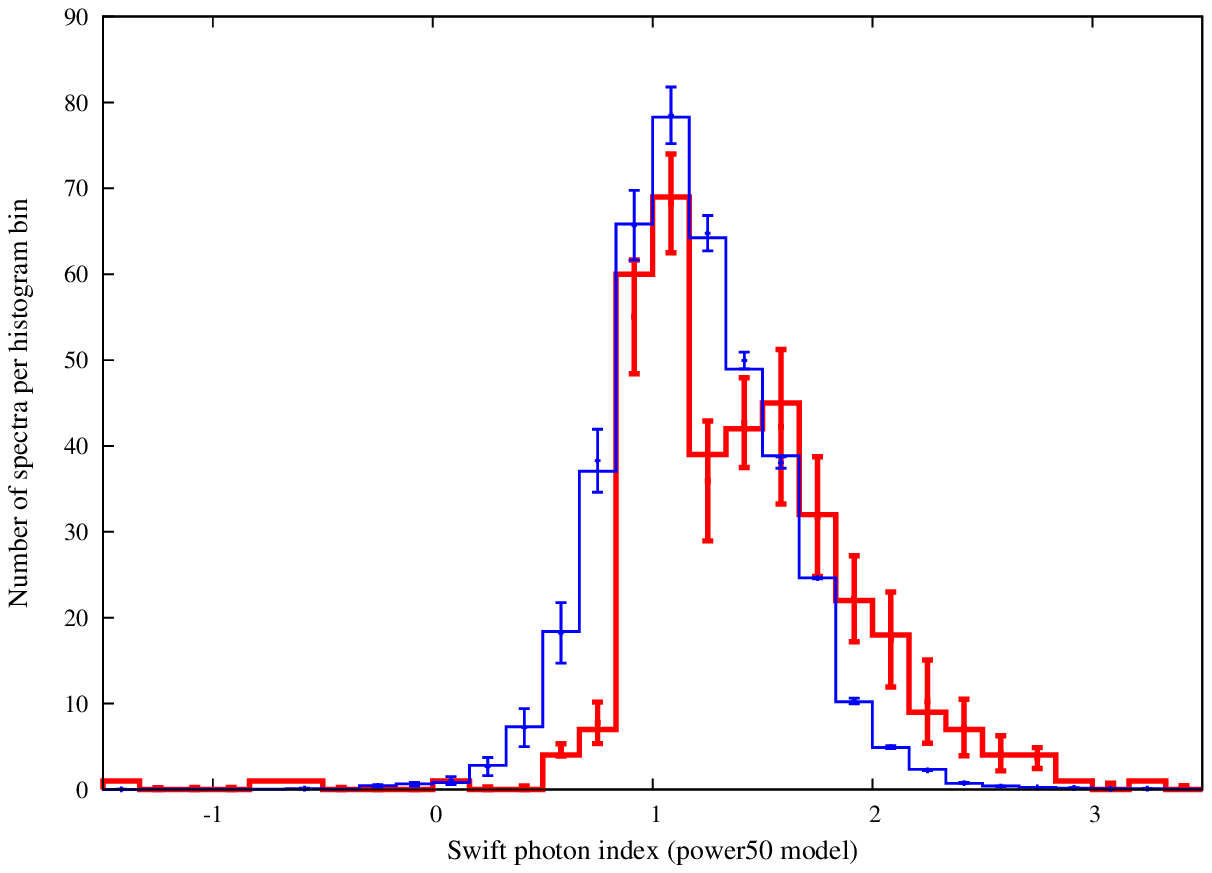}
\caption{Distribution of photon indices of the time resolved spectra of \swift\ GRBs (thick red lines). The upper panel shows the results for the {\tt cutep50} spectral model. The lower panel shows the result for the {\tt power50} model. For comparison, the distribution of photon indices of the simulated BATSE spectra are shown by the thin blue lines. The distributions of the photon indices of the simulated BATSE spectra are re-scaled to be comparable in height to the \swift\ photon index distributions.}
\label{swift_hist_cutep50}
\end{figure}

The dIfference in the selection criteria result in a difference in the distribution of spectral parameters in the sample of time-resolved spectra of GRBs in the \swift\ GRB sample. This is illustrated in Fig. \ref{swift_hist_cutep50}, where we show the histograms of distributions of the photon indices found in fitting the burst spectra on the cutoff powerlaw and on the simple powerlaw models. Uncertainties on the numbers of photon indices in each bin of the histograms were computed in the same way in the case of Fig. \ref{batse_sw_sim_apha},  via Monte-Carlo simulations of sets of photon indices in which the photon indices are randomized within the errors of measurements, around the value observed in the real data. 

One can see that, as expected, no statistically significant excess of the very hard spectra with indices below $\Gamma=2/3$ is observed. The numbers of spectra with photon indices below $2/3$ and their statistical uncertainties are given in the last two rows of the Table \ref{tnumbbeyond} for the two spectral models used in our analysis.

For comparison, we also show in Fig. \ref{swift_hist_cutep50} the histograms of distributions of the photon indices found from  simulations of the BATSE GRB spectra, discussed in Section \ref{simulations}. One can see that slightly different choices of selection criteria for the \swift\ GRBs result in the appearance of larger number of softer time resolved spectra in \swift\ sample, compared to the BATSE sample. This is readily explained by the fact that in the included weaker GRBs longer integration times are needed to achieve the required signal to noise ratio. The GRBs typically exhibit the so-called hard-to-soft spectral evolution. This implies that with longer integration time one finds, in general, a softer spectrum.

\section{ Discussion }\label{discuss}

The goal of our study of time-resolved spectral of \swift\ GRBs was to clarify if \swift\ is able to confirm or disprove BATSE observation that some $\sim 30$\% of spectra of the prompt emission phase are characterized by very hard low-energy photon indices, $\Gamma<2/3$ \citep{preece98a}. This observation is potentially important because it rules out the optically thin synchrotron emission model of the prompt emission phase. 

 The validity of the  BATSE result was questioned by by non-observation of the excessively hard GRB spectra with HETE \citep{heteII}. It was expected that an independent test of validity of BATSE result will become available with the help of \swift, which is sensitive in the energy range 15-150~keV, optimal for the precision measurement of the low energy spectral indices of the GRB prompt emission phase \citep{BB04}.

Our study shows that
\begin{enumerate}
\item it is not possible to fully test the BATSE claim of existence of too hard GRB spectra "beyond the synchrotron deathline" with \swift\ and
\item the statistics of the spectral parameters of \swift\ GRBs is consistent with the assumption of existence of a significant fraction of prompt emission spectra beyond the "synchrotron deathline".
\end{enumerate}

Contrary to the initial expectations, the constraints on the low-energy photon indices, found from the \swift\ data are much weaker than the ones found from the BATSE data. This is explained by several differences between BATSE telescope on board of {\it CGRO} and BAT instrument on board of \swift.

The limited energy range of BAT instrument (sensitive at the energies up to 150~keV) precludes a measurement of the cut-off or break energies in a large fraction of time-resolved GRB spectra. The uncertainty in the cut-off energy introduces a large uncertainty in the measurement of the photon index, if the GRB spectra are fit with a cut-off powerlaw model ({\tt cutep50}). This effect is illustrated in Fig. \ref{ecut_contour}, showing the error contours in the $E_{\rm cut}-\Gamma$ parameter space for a typical GRB spectrum from a \swift\ GRB sample. Larger measurement errors affect the distribution of photon indices of the time resolved GRB spectra. This leads to  "washing out" of the statistically significant hard tail in the distribution of the photon indices. This is demonstrated in the 4-th column of Table \ref{tnumbbeyond}, where the statistics of the GRB spectra with the photon indices harder than $2/3$ is shown. One can see that even if the size of the sample of the time resolved GRB spectra in \swift\ would be comparable to the one of the sample of BATSE time resolved spectra ($\sim 8\times 10^3$ spectra), it would not be possible to test the existence of the harder than $\Gamma=2/3$ spectra with the \swift\ data, if the cut-off powerlaw model ({\tt cutep50}) would be used for the spectral fitting. 

One can see from Table \ref{tnumbbeyond} that if the size of the \swift\ GRB spectral sample would be comparable to the one of BATSE, a statistical test of the BATSE result on the hard GRB spectra would still be possible if the data are fit with the simple powerlaw model ({\tt power50}) without the high-energy cut-off. Spectral fitting in this model does not suffer from the uncertainty related to the uncertainty of measurement of the cut-off energy, mentioned above. This leads to tighter constraints on the measurement of the photon index. 

However, a ten times smaller field of view and shorter operation time of \swift\ explain the much smaller size of the sample of the time resolved GRB spectra, compared to the BATSE spectra sample. We have been able to generate some $\sim 300$ time resolved GRB spectra, using a selection criterion, similar to the one used in the BATSE time resolved spectral analysis (sufficient signal to noise ratio within a given time interval). This amount is more than a factor of 20 smaller than the number of time resolved spectra in the BATSE database. Lower statistics of the GRB spectral sample further reduces the possibility to test the BATSE result, even if the powerlaw model is used for the spectral fitting.  As one can see from Table \ref{tnumbbeyond}, the expected statistical uncertainty of the numbers of GRB spectra with given photon indices is too large, so that the spectra with $\Gamma<2/3$ still can be easily confused with the statistical scatter of the spectra with $\Gamma>2/3$ within the measurement errors.

Having analyzed the limitations of the BAT instrument via simulations of the observational appearance of BATSE GRBs in the BAT detector, we have found that the only available possibility to test the BATSE result with \swift\ data is to systematically search for the episodes of spectral hardening beyond $\Gamma=2/3$ in the entire \swift\ GRB sample. The idea was to find (or constrain the amount of) the time intervals with relatively high flux and hard spectrum, so that the errors of measurement of the photon index are small enough and the measured photon index is harder than $\Gamma=2/3$ at $\ge 3\sigma$ confidence level. If found, the photon indices such "high confidence" hard spectra can not be confused with the the softer photon indices which occasionally could appear to look harder because of the scatter within the measurement error. An estimate of the expected number of GRBs with such "high confidence" hard spectra in \swift\ shows that some 2 to 5 GRBs with sufficiently bright episodes of hardening should have been detected over $\sim 4$~years of \swift\ mission. 

Our systematic search of the episodes of "high confidence" hardening of \swift\ GRB spectra has revealed that if the time binning of the GRBs is chosen in a way similar to the one adopted for the analysis of BATSE time resolved GRB spectra, only two GRBs have sufficiently bright and long episode of spectral hardening. Remarkably, these are the two highest fluence GRBs in the \swift\ GRB sample. The spectral hardening is observed in 5 time intervals in total, compared to  $\sim 30$ expected from the re-scaling of BATSE data. We have demonstrated that the slight discrepancy between the real and the simulated data disappears if one takes into account the fact that the errors of measurement of the photon index in the simulated data are under-estimated, because the simulation procedure ignores the possibility that GRB could be detected in the partially coded field of view. 

Relaxing the requirement on the signal to noise ratio imposed for the time binning of the GRB lightcurves for the time resolved spectral analysis, we have found that, in fact, more GRBs in the \swift\ GRB sample possess the intervals of hardening beyond $\Gamma=2/3$. The lightcurves of these bursts and the evolution of the photon indices over the burst duration is shown in Fig. \ref{lightcurves}.
Given the very low statistics of the observations of "high confidence" hard spectra in the \swift\ GRB sample, the only firm conclusion which can be drawn at the moment is that \swift\ data confirm the fact of {\it existence} of the episodes of spectral hardening of the GRB prompt emission beyond the "synchrotron deathline".

To summarize the results of our investigation, we find that statistics of detections of the episodes of spectral hardening of the GRB spectra beyond $\Gamma=2/3$ in \swift\ is consistent with previous BATSE result on the existence of a significant fraction of time resolved GRB spectra "beyond the synchrotron deathline". 

The existence of the GRB spectra "beyond the synchrotron deathline" rules out a model in which prompt $\gamma$-ray emission of GRB is the optically  thin synchrotron emission from a population of electrons moving in a random magnetic field (the model most commonly adopted within the "relativistic fireball" approach for the modeling of prompt emission)  \citep{ZM04,Piran05}. Several possible solutions of the problem of the "too hard" photon indices are available, such as the synchrotron/curvature/"jitter" emission mechanism by electrons are moving at small pitch angles $\theta\sim \gamma^{-1}$ in magnetic field \citep{epstein73,BB04,lloyd02,medvedev00}, or  optically thick synchrotron mechanism with the synchrotron self-absorption energy reaching $\sim 100$~keV \citep{lloyd00}, or the inverse Compton scattering mechanism \citep{panaitescu00,kumar,piran08,shemi94,shaviv95,lazatti00,DR04}.

\section*{ Acknowledgments }

We would like to thank T.Courvoisier and V.Beckmann for the useful discussions of the subject and for the comments on the manuscript.

\bibliographystyle{plain}

 \label{lastpage}

\end{document}